\documentclass[twocolumn,preprintnumbers,amsmath,amssymb,prl]{revtex4-1}

\usepackage{amsmath,amssymb,epsfig,color}

\usepackage{dcolumn}

\def\be{\begin{eqnarray}}
\def\ee{\end{eqnarray}}
\def\bea{\begin{eqnarray}}
\def\eea{\end{eqnarray}}

\begin{document}

\preprint{}

\title{
Holomorphy, triality and non-perturbative beta function in 2d supersymmetric QCD
}

\author{Abhijit Gadde}

\affiliation{Institute for Advanced Study, Princeton, NJ 08540
}

\date{\today}

\begin{abstract}
In this paper, we study the RG flow in the non-linear sigma models obtained from a 2d ${\cal N}=(0,2)$ supersymmetric QCD. The sigma model is parameterized by a single Kahler modulus. We determine its exact non-perturbative beta function using holomorphy, triality and the knowledge of the infra-red fixed point.
\end{abstract}


\maketitle


\noindent{\bf Picture of the RG  flow in $(0,2)$ gauge theories:\;\;\;}
A distinctive feature of the two dimensional ${\cal N}=(0,2)$ supersymmetric  theories is that they are chiral. As a result its global symmetries generically have non-vanishing 't Hooft anomalies. The anomaly matching argument suggests that the low energy theory consists of the gapless modes (that contribute to the 't Hooft anomalies in the infra-red). Their autonomous physics is described by a conformal field theory. On the other hand, the gauge coupling in two dimensions is a relevant deformation with mass dimension~$1$. These two statements together imply that $(0,2)$ gauge theories admit non-trivial renormalization group flows to conformal fixed points. In fact, in most cases the RG flow naturally splits into two stages. In the first stage, the gauge coupling and other dimensionful parameters flow rapidly to infinity. A good description towards the end of this phase is in terms of a non-linear sigma model. The second stage of the RG flow takes place in the K\"ahler and complex structure moduli space of the sigma model. Because these moduli are classically dimensionless, the RG flow is logarithmic at one loop. Eventually it takes the theory to a conformal fixed point. In certain special cases, the conformal field theory may have exactly marginal directions but we will not study such examples in this paper.

In \cite{Gadde:2013lxa}, it was discovered that a large class of $(0,2)$ theories exhibit low energy dualities similar to the Seiberg duality of the four dimensional ${\cal N}=1$ supersymmetric theories. Moreover, the end point of their RG flow was identified in \cite{Gadde:2014ppa}, in terms of an explicit conformal field theory. In this paper, we will obtain an exact non-perturbative description of the second stage of this RG flow i.e. of the RG flow in the associated non-linear sigma model. 

A prototypical theory belonging to this class is the $(0,2)$ supersymmetric QCD: A $U(N_c)$ gauge theory coupled to fundamental matter multiplets. We summarize the matter field content and their transformation properties under gauge and global symmetries in table \ref{content}.
\begin{table}[h]
\be
\begin{array}{@{~~}c@{~~}|@{~~}c@{~~~}c@{~~~}c@{~~~}c@{~~}|@{\quad}c@{~~}l@{~~}}
 & \Phi & \Psi & P & \Gamma & & \text{labels} \\
\hline
U(N_{c}) & \square & \overline{\square} & \overline{\square} & {\bf 1} & & \alpha,\beta,\gamma \tabularnewline
SU(N_{1}) & {\bf 1} & {\bf 1} & \square & \overline{\square} & & a,\,b,\,c  \tabularnewline
SU(N_{2}) & \overline{\square} & {\bf 1} & {\bf 1} & \square & & r,\,\,s,\,t \tabularnewline
SU(N_{3}) & {\bf 1} & \square & {\bf 1} & {\bf 1} & & i\,\,,j,\,k
\end{array}
\label{reps}\nonumber
\ee
\caption{Field content of the $(0,2)$ SQCD.}\label{content}
\end{table}
Here $\Phi$ and $P$ are chiral multiplets while $\Psi$ and $\Gamma$ are fermi multiplets.  The cancellation of the $SU(N_c)$ gauge anomaly determines $N_c=(N_1+N_2-N_3)/2$. In order to cancel the anomaly of the $U(1)$ factor, we also add two fermi multiplets $\Omega$ with $+1$ charge under the gauge group $U(1)$.

In addition to the gauge interaction, the theory has a holomorphic J-term superpotential 
$$
m \int d\theta\, \Gamma^s_a \Phi_\alpha^s P^\alpha_a+ t\int d\theta\, {\rm tr}\, \Lambda.
$$
Here  $\Lambda$ is the gaugino fermi multiplet. The first term  leads to a Yukawa interaction as well as a quartic interaction for scalar fields and the second term is the complexified Fayet-Illiopolous (FI) coupling. Taking $t:=i\zeta+\theta/2\pi$, the FI coupling appears in the component  Lagrangian as  
$$
\zeta \int d^2x \,  {\rm tr}\,D+\frac{\theta}{2\pi}  \int {\rm tr}\,F.
$$
Due to the periodicity of the $\theta$ angle, it is more convenient to use the exponentiated variable $z:=e^{2\pi i t}$. We label this theory ${\cal T}_{\rm gauge}[g_{\rm YM}, m, z]$.  

The most convenient description of this $(0,2)$ supersymmetric QCD depends on the energy scale at which it is being studied. 
The couplings $g_{\rm YM}$ and $m$ have mass dimension $1$ while $t$ is classically marginal. As pointed out earlier, during the first stage of the renormalization group flow, the classically relevant couplings flow rapidly to infinity. By the end of this flow, the convenient description of the theory is in terms of a non-linear sigma model.  The holomorphic parameter $z$ is the exponentiated K\"ahler modulus of the sigma model. As we will see shortly, this the only modulus of the sigma model.
The next phase of the RG flow takes place in this space. Let us denote the point at which the flow terminates as $z_*$. This is the conformal fixed point identified in \cite{Gadde:2014ppa}. The picture of the RG flow is
$$
{\cal T}_{\rm gauge}[g,m,z] \xrightarrow[g_{\rm YM},m\to \infty] {\quad \text{RG flow} \quad}{\cal T}[z] \xrightarrow[z\to z_*]{\quad \text{RG flow} \quad} {\cal T}_{\rm CFT}.
$$
We have used the notation  ${\cal T}[z]$ to denote the sigma model with the exponentiated K\"ahler modulus $z$. 
In this note we will be studying the second phase of the RG flow in a holomorphic renormalization scheme i.e. in a scheme that respects the holomorphy of the Kahler modulus.

In \cite{Gadde:2013lxa}, it was suggested, using the superconformal index, that the supersymmetry is broken  unless $N_1,N_2$ and $N_3$ satisfy triangle inequality. In what follows, until otherwise mentioned, we will assume that $N_1,N_2$ and $N_3$ do obey triangle inequality and hence that the supersymmetry  is preserved.
 
The target space of the $(0,2)$ non-linear sigma model is a holomorphic vector bundle $E\to M$ over a K\"ahler manifold $M$. Anomaly cancellation requires $ch_2(E)=ch_2(TM)$. For the case at hand, the target space of the sigma model is the vacuum manifold of the gauge theory. It is obtained by solving the $D$-term and``$J$-term" constraints modulo gauge symmetry action.
\begin{eqnarray}
P^a_\alpha {\bar P}_a^\beta -\Phi^\beta_s {\bar \Phi}_\alpha^s-\zeta \delta_\alpha^\beta&=&0 \nonumber\\
P^a_\alpha \Phi^\alpha_s &=& 0\nonumber
\end{eqnarray}
They imply $\Phi =0$ (resp. $P$=0) for $\zeta >0$ (resp. $\zeta<0$). Dividing by the $U (N_c)$ gauge group, we get the space $Gr(N_c,N_1)$.
The Fermi fields engineer fibers of the holomorphic vector bundle.
As the field $\Psi$ transforms in the fundamental representation of the gauge group,
it forms a fiber of the universal subbundle (tautological bundle) $S$.
The field $\Gamma$ is neutral but it satisfies the $J$-term relation
\be
\Gamma_a^s P^a_\alpha=0. \nonumber
\ee
Therefore, $\Gamma$ furnish a fiber of the universal quotient bundle (orthogonal bundle) $Q$, which is defined through the short exact sequence:
\be
0\longrightarrow S \longrightarrow {\cal O}^{N_1} \longrightarrow Q \longrightarrow0. \nonumber
\ee
 All in all, for $\zeta >0$, the gauge theory ${\cal T}_{\rm guage}$ flows to the nonlinear sigma model with the target space,
\be\label{sigma-model1}
S^{\oplus N_3} \oplus Q^{\oplus N_2} \longrightarrow Gr(N_c,N_1).
\ee
For $\zeta <0$, the $D$-term equation gives vev only to $\Phi$.
Similar arguments lead to the target space $S^{*\oplus N_3} \oplus Q^{*\oplus N_1} \longrightarrow Gr(N_c,N_2)$.
\\

\noindent{\bf Charge conjugation:\;\;\;}
Theories with $(0,2)$ supersymmetry are generically not charge conjugation invariant. The action of charge conjugation on ${\cal T}_{\rm gauge}$ replaces representations of all the fields by their  complex conjugates. In the resulting sigma model, $N_1$ and $N_2$ are exchanged and the fibers of the holomorphic vector bundle are complex conjugated. From the discussion in the previous section, we see that this is the same sigma model that is obtained by changing the sign of $\zeta$.  Analogous to the spurion analysis in four dimensional ${\cal N}=1$ supersymmetric theories, the complexified FI parameter $t$ can be thought of as the background value of a chiral superfield. All the transformations therefore should be written in a way that preserves the holomorphy of $t$.
This implies that the charge conjugation ${\cal C}$ is implemented at the level of the sigma model as $z\to1/z$. 
\be\label{conjz}
{\cal C}\cdot{\cal T}[z]={\cal T}[1/z].
\ee
As a result the sigma model physics should be invariant under the simultaneous exchange $N_1 \leftrightarrow N_2, z\leftrightarrow 1/z$.

In \cite{Gadde:2014ppa}, it was proposed that the microscopic theory flows to one of the two possible conformal fixed points that are related to each other by charge conjugation. Because the renormalization group flow commutes with charge conjugation, it follows that  if the theory ${\cal T}[z]$ flows to the fixed point at $z_*$ then the theory ${\cal T}[1/z]$ should flow to the charge conjugate fixed point. Moreover, from equation \eqref{conjz} we see that this other fixed point is at $1/z_*$. This discussion implies that the FI parameter space is divided into two regions according to their attractors and the regions are  mapped into each other by $z\to 1/z$. 
\\

\noindent{\bf Consequence of Triality:\;\;\;}
Let ${\cal T}'$ and ${\cal T}''$ be the theories obtained from ${\cal T}$ by cyclic permutation of $\{N_1, N_2,N_3\}$. We denote their labels by $z'$ and $z''$ respectively. 
It was argued in \cite{Gadde:2013lxa} that all three theories flow to the same fixed point (modulo charge conjugation) i.e. we have the isomorphism
\be
{\cal T}[z_*]\cong {\cal T}'[z'_*]\cong {\cal T}''[z''_*]. \nonumber
\ee 
Moreover, thanks to the equivalence relations, 
\bea
S\to Gr(k,n) \,&\qquad \sim \qquad & Q^* \to Gr(n-k,n), \nonumber\\
Q \to Gr(k,n) & \qquad \sim \qquad & \,S^*\to Gr(n-k,n).\nonumber
\eea
we have further isomorphisms
\be
{\cal T}[\infty]\cong{\cal T}'[0],\quad {\cal T}'[\infty]\cong{\cal T}''[0],\quad {\cal T}''[\infty]\cong{\cal T}[0]. \nonumber
\ee
This identification effectively glues together the K\"ahler moduli spaces of all three Grassmannians.
Alternatively, it means that in the K\"ahler moduli space of the sigma model there are three points with distinct large volume descriptions. This seems to be a novel phenomenon.

\begin{figure}[h]
\centering
\includegraphics[scale=0.3]{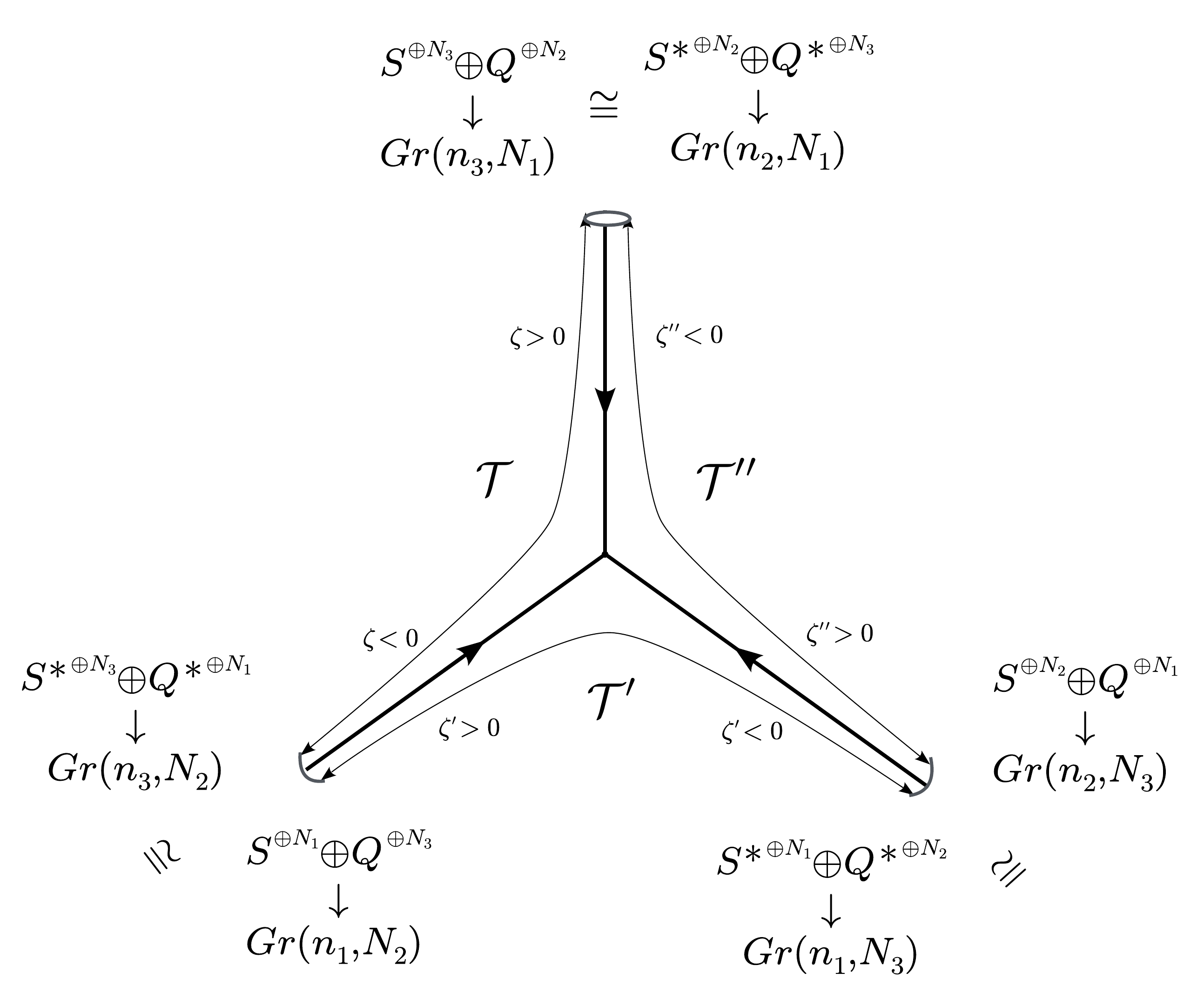}
\caption{Parameter space of the exponentiated FI parameter $z$ and it's various limits. The arrows denote the qualitative expectation for the RG flow with a similar set of arrows on the ``back side".}
\label{FIspace}
\end{figure}

The  moduli space admits three patches of coordinates $\{z,z',z''\}$ useful for describing each of the three pairs.
This leads to a qualitative picture of  the K\"ahler moduli space drawn in figure ~\ref{FIspace}.
Consider the transition map $f(z)$ between different coordinate patches such that ${\cal T}[z]\cong {\cal T}'[f(z)]\cong{\cal T}''[f^2(z)]$  for all points in the $z$ plane. The complexified Fayet-Illiopolous parameter is the background value of a non-dynamical chiral multiplet, we expect the map to respect its holomorphy.  We also expect the map to be bijective with the property $f^3(z)=z$.  The only function (up to coordinate normalization) having all the desired properties is
\be\label{trialityz}
f(z)=\frac{1}{1-z}. 
\ee
Here we have normalized the coordinates such that ${\cal T}'[\infty]\cong{\cal T}[1]$ and so on. The ambiguity in the normalization can be restored by substituting $z$ by $z/z_0$. We will take $z_0=1$ to avoid the clutter. We will see later that it is indeed the correct normalization.

The physics of the sigma model should be consistent with the triality transformation. In particular,  the flow equation should be invariant under the simultaneous operation $z\to f(z)$ and $N_1\to N_2\to N_3\to N_1$.
\\

\begin{figure}[h]
\centering
\includegraphics[scale=0.3]{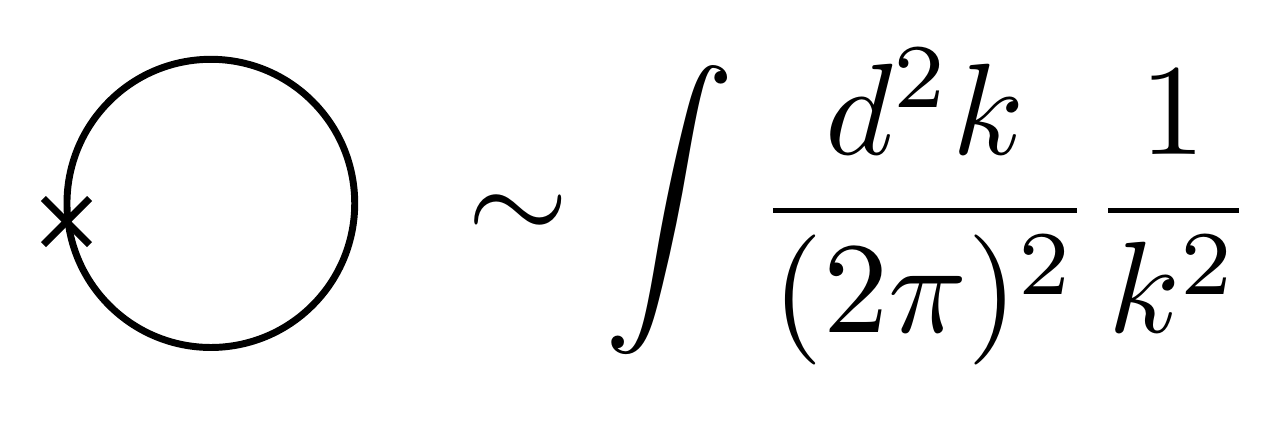}
\caption{Running of the FI parameter at one-loop in $g_{\rm YM}$.}
\label{tadpolebeta}
\end{figure}

\noindent{\bf Holomorphic beta function:\;\;\;}
The Fayet-Illiopolous parameter $t$ is classically marginal but it does run quantum mechanically. At leading order, its  beta function receives a contribution at one-loop via the tadpole diagram, shown in figure \ref{tadpolebeta}. 
The D term only couples to the scale fields. For $\zeta>0$, $P$ gets a vacuum expectation value. Due to this and the J-term interaction $m\int \Gamma \Phi P$, the scalar field $\Phi$ becomes parametrically massive. As a result, the contribution of $\Phi$ to the loop integral are parametrically suppressed. The only fields running in the loop are $P$. The leading order beta function is
\be\label{one-loop}
\frac{dt}{d\log \Lambda}=\frac{N_1}{2\pi}.
\ee
The same leading order result is obtained from the sigma model perturbation theory. This confirms our argument about $\Phi$ becoming parametrically massive in the supersymmetric vacuum. 

The beta function does receive contribution due to non-perturbative effects in $t$. These are the so called world-sheet instantons in the sigma model or equivalently,  the vortex configurations in the gauge theory. The non-perturbative contributions vanish in the large volume limit of the sigma model i.e. in the limit $\zeta\to \infty$ ($z\to 0$). Then the one-loop beta function \eqref{one-loop} is a good approximation.  It is convenient to rephrase the perturbative beta function as a one-form valued constraint on the exact holomorphic beta function
\be\label{zto0}
i\,d\log \Lambda \xrightarrow{z\to0} \frac{1}{N_1}\frac{dz}{z}.
\ee
A similar analysis in the $z\to \infty$ limit yields
\be\label{ztoinfty}
i\,d\log \Lambda \xrightarrow{z\to \infty}-\frac{1}{N_2}\frac{dz}{z}.
\ee
Indeed the limits are  invariant under the simultaneous exchange $N_1,\leftrightarrow N_2, z \leftrightarrow1/z$ as expected.
Recall that $z\to\infty$ limit of theory ${\cal T}$ is identical to $z'\to~0$ limit of theory ${\cal T}'$. Moreover, $z'\to \infty$ limit of theory ${\cal T}'$ is identical to $z\to 1$ limit of theory ${\cal T}$. In this limit, the one loop beta function, computed in the $z'$ variable, follows the same analysis as before.
\be\label{zptoinfty}
i\,d\log \Lambda \xrightarrow{z'\to \infty}-\frac{1}{N_3}\frac{dz'}{z'}.
\ee
Using the triality transformation \eqref{trialityz}, we can rewrite this as
\be\label{zto1}
i\,d\log \Lambda \xrightarrow{z\to 1}\frac{1}{N_3}\frac{dz}{z-1}.
\ee
The three limits, equations \eqref{zto0},\eqref{ztoinfty} and \eqref{zto1} serve as the basis for fixing the exact holomorphic beta function.
We see that the differential $i\, d\log \Lambda$ has poles at $z=0, \infty$ and $1$ with residues $\frac1N_1,\frac1N_2$ and $\frac1N_3$ respectively.
The poles of the $i\,d\log\Lambda$ are the zeroes of the beta function. The ones  listed above correspond to the large volume limits of the sigma model and hence to the UV fixed points for the RG flow of  $z$. As discussed earlier, in addition to the three UV fixed points, the only other fixed points are the two  infra-red fixed points. They are at $z_*$ and $1/z_*$. Because the total residue on the complex plane has to be zero, the sum of the residues at the IR fixed points must be $-\sum_{i=1}^3\frac1N_i$. The  exact  beta function has the form,
\bea
&&i\, d\log \Lambda = \frac{1}{N_1}\frac{dz}{z}+\frac{1}{N_3}\frac{dz}{z-1}\\
&&-\Big(\frac12\sum_{i=1}^{3}\frac1N_i+a\Big)\frac{dz}{z-z_*}-\Big(\frac12\sum_{i=1}^{3}\frac1N_i-a\Big)\frac{dz}{z-1/z_*}\nonumber \\
&&+ {\, \rm possible\, higher \, order\, singularities\, at}\, z_*{\rm \,and}\, 1/z_*. \nonumber
\eea

We can restore the ambiguity in the normalization by substituting $z$ by $z/z_0$. Using the invariance of the beta function under $z\to 1/z, N_1\leftrightarrow N_2$, we see that $z_0$ has to be $1$. It also determines $a=0$. The infra-red fixed point $z_*$ is fixed by requiring consistency with triality i.e. invariance under the simultaneous operation $z\to f(z)$ and $N_1\to N_2\to N_3\to N_1$.
We see that it has to be the fixed point of the map $f(z)$. Solving $z_*=f(z_*)$ we get $z_*=e^{\frac{i\pi}{3}}$. Happily the other fixed point is $1/z_*$ as desired. Together, these two constraints also imply that the higher order singularities are absent. Although we have checked these facts for some special ansatz, it would be nice to obtain a rigorous proof.
With these conditions the exact  beta function becomes
\be\label{exactbeta}
\boxed{
\frac{dz}{d\log \Lambda} = i\Big(\frac{1}{N_1}\frac{1}{z}+\frac{1}{N_3}\frac{1}{z-1}-\sum_{i=1}^{3}\frac1N_i \frac{z-\frac12}{z^2-z+1}\Big)^{-1}.\nonumber
}
\ee
The exact formula can be expanded to read off the perturbative and non-perturbative corrections. 
The non-perturbative configuration with $k$ vortices contributes a term proportional to $z^k$. 
\be
\frac{dt}{d\log\Lambda}=\frac{N_1}{2\pi}-\frac{1}{4\pi}\Big(N_1+\frac{N_1}{N_2}-\frac{N_1}{N_3}\Big)z+\ldots.
\ee
We see that there are no higher-loop perturbative corrections.

\begin{figure}[h]
\centering
\includegraphics[scale=0.4]{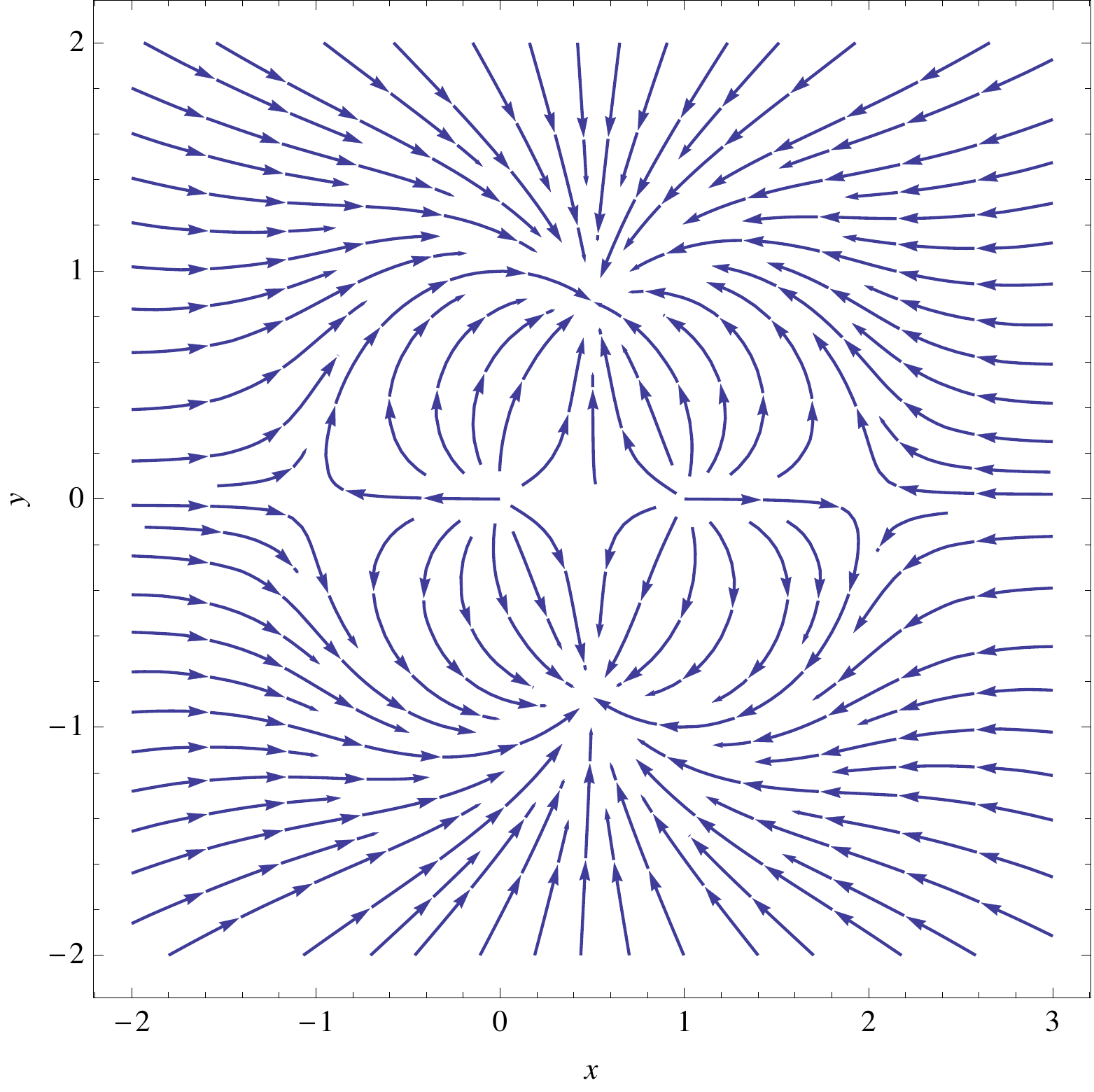}
\caption{Flow lines of the renormalization group of $z=x+iy$.}
\label{flow-lines}
\end{figure}

We have plotted the flow lines of the renormalization group for $N_1=N_2=N_3$ in figure \ref{flow-lines}. They agree with our qualitative expectation from figure \ref{FIspace}.

\noindent{\bf Theories dual to free fermions:\;\;\;}
Let us analyze the case when one of the triangle inequalities of $\{N_1,N_2,N_3\}$ is saturated, say $N_2+N_3=N_1$. Then theory ${\cal T}'$ has a trivial gauge group and thus consists of only free Fermi multiplets. Naturally, there is no FI parameter $\zeta'$ associated with it. However, the $\zeta\to0$ limit of the theory $\cal T$ is still a Grassmannian non-linear sigma model. In fact, that is the only large volume point in its K\"ahler moduli space. This changes the previous analysis. Now we have only one UV fixed point and only one IR fixed point i.e. free left-moving fermions. The residue of the pole of $i\,d\log\Lambda$ at $z\to0$ is $1/N_1$. The residue at the only other pole, at $z\to \infty$,  has to be $-1/N_1$. Thus the exact beta function is
\be
\frac{dz}{d\log \Lambda}=i\,N_1 z.
\ee
Curiously, it seems that the beta function does not receive any non-perturbative corrections.

In the case where the triangle inequality of $N_i$'s is not obeyed the supersymmetry is broken. The IR fixed point theory or even the number of possible IR fixed points is not known. This prevents us from carrying out a similar analysis in this case.

\noindent{\bf Discussion:\;\;\;}
In two dimensions the renormalization group flow is a gradient flow with respect to the Zamolodchikov c-function
\be\label{RGgradient}
\frac{ds_i}{d\log\Lambda}=g_{ij}\frac{\partial c}{\partial s_j}.
\ee
Although, this was proved in conformal perturbation theory in  \cite{Zamolodchikov:1986gt}, compelling arguments for its non-perturbative validity were given in \cite{Friedan:2009ik}. In principle, we could use the gradient flow to learn about the c-function of the theories at hand.

 In the case studied here, there is only a single complex coupling constant $z$. 
We have computed its exact beta function i.e. the left-hand side of the equation \eqref{RGgradient}. 
For the equation to respect the holomorphy of $z$ we expect the metric on the coupling space to be K\"ahler. In the case of $(2,2)$ theories, it was conjectured \cite{Jockers:2012dk} and subsequently proved \cite{Gomis:2012wy} that the K\"ahler metric on the K\"ahler moduli space can be determined from the partition function of the theory on $S^2$. It would be interesting to see if one can similarly localize the path integral of the $(0,2)$ theory to compute the metric. 
In the absence of such technology, we can hope to fix the K\"ahler metric $g_{z\bar z}$ from the knowledge of the singularities of the moduli space. It is an old result that the constant curvature hyperbolic metric on the sphere is uniquely determined from the deficit angles at the marked points \cite{Picard}. Possibly, the deficit angles could follow from the value of the central charge $c$ at the UV and the IR fixed points \cite{Gadde1}.

\noindent{\bf Acknowledgments}:~
We would like to thank Chris Beem, Sergei Gukov, Ilarion Melnikov, Pavel Putrov, Nathan Seiberg, Brian Willett and Edward Witten. This work is supported by the Raymond and Beverly Sackler Foundation Fund and the NSF grant PHY-1314311.

\end{document}